\documentclass[aps,rmp,floats,superscriptaddress,twocolumn,final]{revtex4-2}
\usepackage[utf8]{inputenc}

\usepackage{graphicx,dcolumn,bm,epic,eepic,fleqn,float}
\usepackage{amssymb,amsmath,amsfonts,multirow,rotate,color}
\usepackage{soul,xcolor,url}
\usepackage{float}

\usepackage[percent]{overpic}
\usepackage{multirow}
\usepackage{booktabs}
\usepackage{siunitx}

\usepackage[hidelinks]{hyperref}
\hypersetup{colorlinks=false,
    linkcolor=black,
    citecolor=black,
    urlcolor=black,
}
\usepackage{xurl}
\usepackage{orcidlink}

\usepackage[capitalise]{cleveref}
\crefname{section}{Sec.}{Sections}

\definecolor{maroon}{rgb}{0.5, 0.0, 0.0}

\setcitestyle{super}

\newcommand{\jaccardindex}[2]{\frac{#1 \cap #2}{#1 \cup #2}}
\newcommand{\jaccarddistance}[2]{1 - \jaccardindex{#1}{#2}}

\begin{document} 

\title{Mapping robust multiscale communities in chromosome contact networks}

\author{Anton Holmgren \orcidlink{0000-0001-5859-4073}}
\email{anton.holmgren@umu.se}
\affiliation{Integrated Science Lab, Department of Physics, Ume\aa~University, Ume\aa, Sweden}
\author{Dolores Bernenko \orcidlink{0000-0002-6618-8232}
}
\email{dolores.bernenko@umu.se}
\affiliation{Integrated Science Lab, Department of Physics, Ume\aa~University, Ume\aa, Sweden}
\author{Ludvig Lizana \orcidlink{0000-0003-3174-8145}}
\email{ludvig.lizana@umu.se}
\affiliation{Integrated Science Lab, Department of Physics, Ume\aa~University, Ume\aa, Sweden}

\date{\today}

\begin{abstract}
To better understand DNA's 3D folding in cell nuclei, researchers developed chromosome capture methods such as Hi-C that measure the contact frequencies between all DNA segment pairs across the genome.
As Hi-C data sets often are massive, it is common to use bioinformatics methods to group DNA segments into 3D regions with correlated contact patterns, such as Topologically Associated Domains (TADs) and A/B compartments.
Recently, another research direction emerged that treats the Hi-C data as a network of 3D contacts.
In this representation, one can use community detection algorithms from complex network theory that group nodes into tightly connected mesoscale communities.
However, because Hi-C networks are so densely connected, several node partitions may represent feasible solutions to the community detection problem but are indistinguishable unless including other data.
Because this limitation is a fundamental property of the network, this problem persists regardless of the community-finding or data-clustering method.
To help remedy this problem, we developed a method that charts the solution landscape of network partitions in Hi-C data from human cells.
Our approach allows us to scan seamlessly through the scales of the network and determine regimes where we can expect reliable community structures.
We find that some scales are more robust than others and that strong clusters may differ significantly.
Our work highlights that finding a robust community structure hinges on thoughtful algorithm design or method cross-evaluation.
\end{abstract}

\maketitle

\section{Introduction}

Mammalian genomes fold into a network of 3D structures that facilitate and regulate genetic processes such as transcription, DNA repair, and epigenetics.\cite{dixon2016chromatin,schwartz2017three, bonev2016organization,denker2016second}
Most recent discoveries linking genetic processes and genomes' 3D organization derive from chromosome capture methods, such as Hi-C.
Hi-C measures the number of contacts between DNA segment pairs and allows researchers to chart chromosome-wide 3D interaction maps.\cite{lieberman2009comprehensive, sexton2012three, dekker2013exploring}
These maps depict chromosomes as having 3D structures on a broad range of scales: megabase-scale A/B compartments,\cite{lieberman2009comprehensive} sub-compartments (A1, A2, B1, \dots, B4),\cite{rao20143d} sub-megabase-scale Topologically Associated Domains (TADs),\cite{dixon2012topological} sub-TADs and short-ranged loops.\cite{rao20143d}
Some of these structures are associated with epigenetic marks, active genes, and chromatin remodelers, such as CCCTC-binding factors (CTCF), cohesin complexes, and CP190.\cite{dixon2012topological, kaushal2021ctcf, remeseiro2016gene, szabo2019principles}
  
Numerous research groups developed methods rooted in bioinformatics to detect significant 3D structures, foremost TADs and A/B compartments.\cite{mackay2020computational, fraser2015hierarchical, liu2021systematic}
However, recently, there has been an emerging research direction alongside this development that takes advantage of the methods developed in complex network theory.
This approach treats the Hi-C data as a weighted network of 3D contacts and groups nodes with above-average connectivity into mesoscale communities.\cite{sarnataro2017structure, lee2019mapping, bernenko2022mapping, boulos2013revealing}
While these and many other community detection methods led to several impactful insights, underneath this approach reside an often overlooked fundamental limitation: in most networks, more than one node partition may represent a feasible network community division.
Because this limitation is fundamental to the network, this type of degeneracy exists regardless of the community-finding method.
Also, the degeneracy becomes increasingly problematic if trying to detect small-scale communities, where there is a significant risk of over-fitting, or in dense networks, where it is hard to determine node-community memberships with significant certainty.\cite{good2010performance} 

This degeneracy problem posits that Hi-C maps' community structure is particularly challenging because Hi-C networks are almost fully connected even if most links are weak.
Therefore, we expect that these networks possess several community divisions that cannot be further rated without including new data, e.g., gene expression or epigenetic profiles.
Yet more intriguing, this limitation hints that there is a noteworthy probability that community-finding or data-clustering algorithms disagree on the optimal division.
This problem likely fueled some debates regarding actual differences between TADs and sub-TADs.\cite{dixon2016chromatin,eres2021tad}

This paper explores these limitations by mapping out the landscape of possible network partitions in Hi-C data.
To this end, we use the Generalized Louvain Method\cite{blondel2008fast, jeubgeneralized}
that allows us to detect communities at different network scales.
We also developed a method to determine regimes where the solution landscape is degenerate and where we find robust communities.

\section{Results}

To study the multiscale 3D organization in chromosomes, we use Hi-C data from the B-lymphoblastoid human cell line (see \cref{sec:data} for references and data handling).
As in other approaches,\cite{bernenko2022mapping, lee2019mapping, sarnataro2017structure} we convert the Hi-C data into a network, where nodes represent $10^5$ base pair long DNA segments (100 kb), and the links stand for segment-segment 3D interactions, where the weights are associated with the Hi-C contact count.
In this study, we focus on chromosome 10.

To partition the network and map out multiscale communities, we use the Generalized Louvain method (GenLouvain).
GenLouvain separates the network into communities where nodes share more interconnections than some null model (we defer details to \cref{sec:louvain}).
To construct a realistic null model, we assume that the segment-segment contact frequencies decay as a power-law $l^{-\alpha}$, with linear separation $l$ and decay exponent $\alpha$.
This scaling feature appears in established polymer physics models\cite{mirny2011fractal}
and in Hi-C data.\cite{pigolotti2020bifractal} 
Averaging the Hi-C contacts over many segments gives two regimes: $\alpha \approx 1.08$ for long distances ($\sim 500$--7000 kb),\cite{rao20143d, lieberman2009comprehensive} and $\alpha \approx 0.75$ for short distances ($\sim 200$--1200 kb).\cite{sanborn2015chromatin}
See \cref{eq:modularity} in \cref{sec:louvain} for how we implement this contact scaling in GenLouvain.

Besides the exponent $\alpha$, GenLouvain has a scale parameter~$\gamma$.
By varying this parameter, users may scan the network hierarchies and find multiscale communities.
Using this approach, we sample feasible partitions of the network.
We call the collection of these partitions the solution landscape.

\subsection{Classifying the solution landscape}

\begin{figure}[htp!]
    \includegraphics[width=0.8\linewidth]{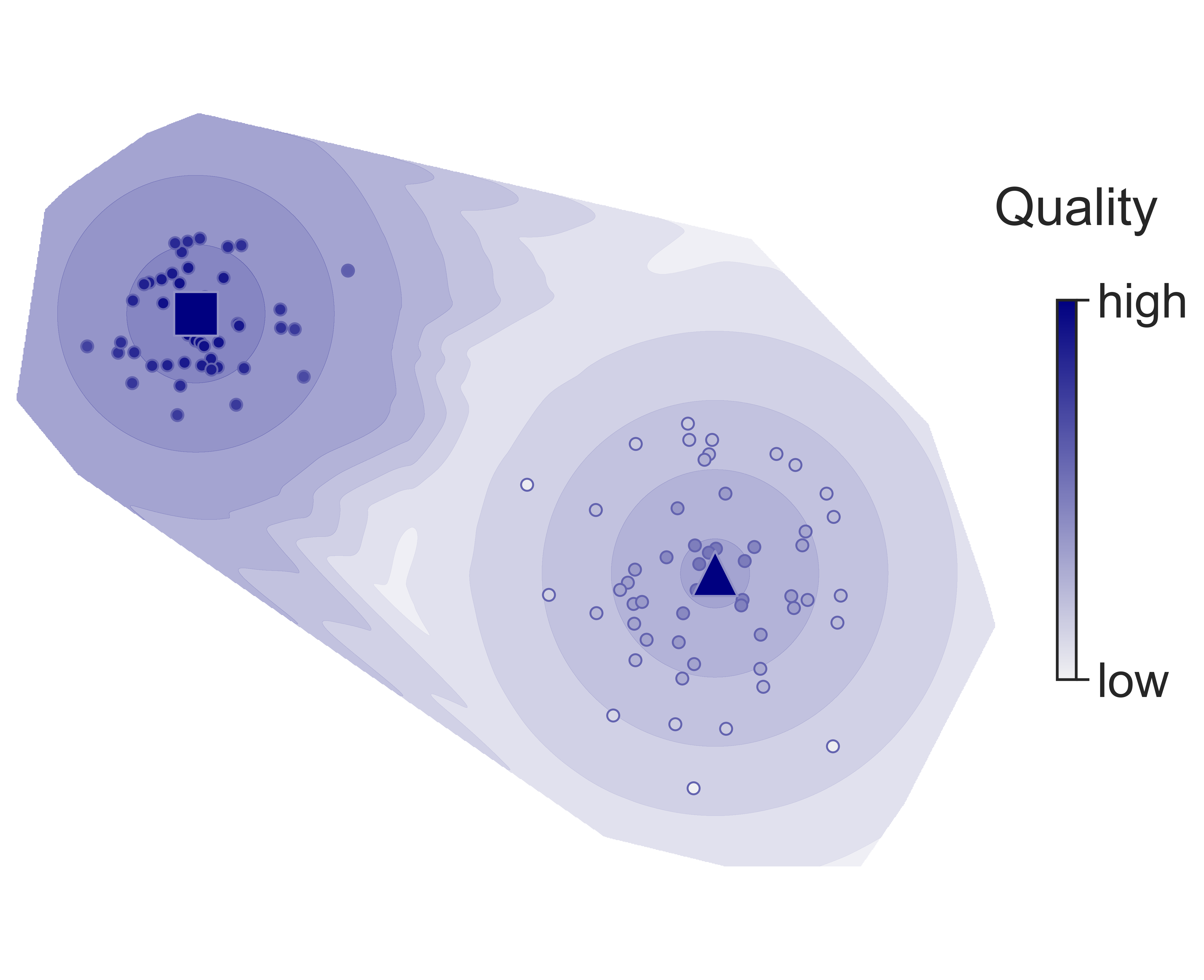}
    \caption{Solution landscape of network partitions (circles) on a quality contour plot. The (locally) best-quality partitions appear on the landscape's peaks ({\small$\blacksquare$} and $\blacktriangle$). The square partition has the highest quality.}
    \label{fig:schematic_sl}
\end{figure}
GenLouvain optimises the modularity quality function~$Q$ (\cref{eq:modularity}) to find mesoscale communities with above-average connectivity.
Because the community division problem is NP-hard, it is practically impossible to enumerate all network divisions and determine which one is optimal.
Instead, GenLouvain finds feasible divisions using a stochastic search algorithm.\cite{de2011generalized}
But as with most community detection algorithms, GenLouvain sometimes gets trapped in local quality maxima.
We illustrate this trapping schematically in \cref{fig:schematic_sl} that shows two well-separated local maxima, {\small$\blacksquare$} and $\blacktriangle$, overlayed in a quality contour plot.
Depending on starting conditions, GenLouvain will gravitate to $\blacktriangle$ or drift towards {\small$\blacksquare$}.
To increase the chance of finding the best-quality partition, we run 1,000 independent optimisation passes using different random seeds and compare the $Q$ values. 

But for some networks, the solution landscape does not split into two distinct peaks as in \cref{fig:schematic_sl}.
For example, the quality may be nearly identical even in distant parts of this landscape.
This means that it is challenging to distinguish the optimal partition since they are degenerate.
To detect such degeneracies, we calculate the distance between partitions $P$ and $P'$ using the weighted mean Jaccard distance
\begin{equation}
    d_{PP'} = \sum_i \min_j \left( \jaccarddistance{C_i^P}{C_j^{P'}} \right) \frac{|C_i^P|}{\sum_k |C_k^P|},
\end{equation}
where $C_i^P$ are the nodes in community~$i$ in $P$.\cite{calatayud2019exploring} 
Because the distances $d_{PP'}$ are not symmetric ($d_{PP'} \neq d_{P'P}$), we use the average:
\begin{equation}\label{eq:jaccard-distance}
    d = \frac{d_{PP'} + d_{P'P}}{2}.
\end{equation}
When $d=0$, the partitions are identical. And if $d=1$, they are completely dissimilar.
We acknowledge that there are other thinkable distance metrics, such as variation of information, but using such metrics will not change the solution landscape's qualitative topology.\cite{good2010performance}

\begin{table}[htbp]
    \centering
    \caption{Solution landscape classification. Var$(Q)$ and Var$(d)$ denote the variability in partition quality and pairwise distances between cluster centers.}
    \label{tab:sl-classification}
    \small
{\setlength{\tabcolsep}{1em}
\sffamily
\begin{tabular}{@{}cccc@{}}
                            &       & \\
                            &       & \multicolumn{2}{c}{Var$(Q)$}  \\
                            &       & low & high \\ \addlinespace[2pt]
\multirow{2}{*}{Var$(d)$}   & low   & global maximum & --- \\ \addlinespace[2pt]
                            & high  & \color{red} degenerate & local maxima \\
\end{tabular}}
\end{table}
Next, we classify the solution landscape using the Jaccard distances $d$ and the partition qualities $Q$.
We find three broad landscape categories depending on the variability of $d$ and $Q$, Var$(d)$ and Var$(Q)$, summarised in \cref{tab:sl-classification}.
First, if both Var$(d)$ and Var$(Q)$ are low, we find structurally similar partitions of almost the same quality.
Second, we find dissimilar partitions of different qualities when both are high.
For partitions in the third category (arguably the most interesting case), where Var$(d)$ is high and Var$(Q)$ is low, we may find dissimilar partitions having similar quality where no partition should be preferred over any other.
In our notation, this case represents a degenerate solution landscape.
The fourth regime (low Var$(d)$ and high Var$(Q)$) is unsound as we find similar partitions with relatively large quality differences.
This means that as long as we find similar partitions, there is no need to study the variability in $Q$ to guarantee that GenLouvain found the global quality maximum.

\subsection{Identifying robust core communities}

We identified three solution landscapes in the previous section using the variabilities among the partitions' quality and pairwise distances.
However, this only provides a qualitative assessment of the landscape's overall characteristics.
Even when there are distinct peaks, there are always some deviations close to these peaks, where node assignments may differ.
To quantify these differences, we tessellate the solution landscape by clustering the partitions and determining robust node-community assignments in each cluster. 


\begin{figure}[thp!]
    \centering
    \includegraphics[width=\linewidth]{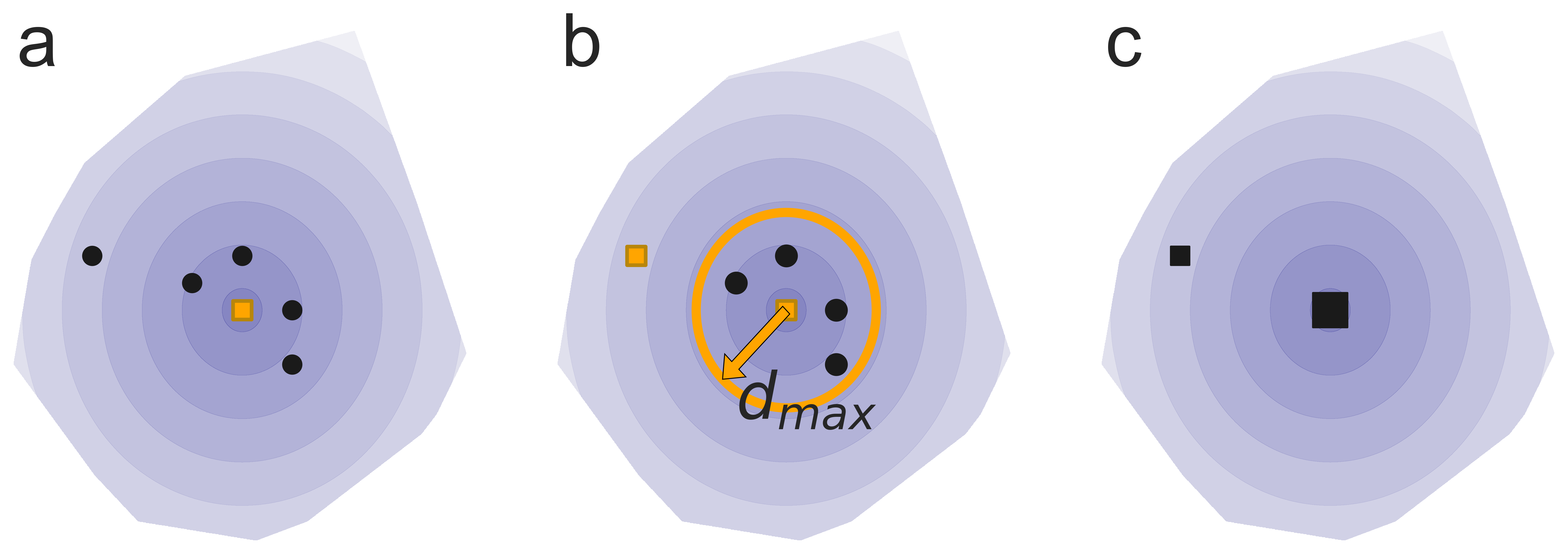}
    \caption{Partition clusters in the solution landscape.
    (a) Partitions with different quality and distance to the best quality partition ({\small\textcolor[HTML]{FAA41C}{$\blacksquare$}}).
    (b) The first partition separated by at least~$d_{max}$ from any cluster centre forms a new centre. This process repeats until all clusters are separated by at least $d_{max}$.
    (c) All partitions are assigned to the nearest cluster.}
    \label{fig:schematic_clustering}
\end{figure}
We start by grouping similar partitions into clusters and comparing their sizes and qualities.
The partition with the locally highest quality represents the cluster centre.
To cluster similar partitions relative to the cluster centre ($d < d_{\rm max}$), we use a clustering algorithm,\cite{calatayud2019exploring} modified to maximise in-cluster similarity.
Below, we summarise the main steps:
\begin{enumerate}
    \item Order all partitions by their quality~$Q$ and let the best partition form a cluster centre (\cref{fig:schematic_clustering}a).
    \item Create new cluster centres with any partitions that are separated by at least $d_{max}$ from any already present cluster centres (\cref{fig:schematic_clustering}b).
    \item Assign the remaining partitions to the closest cluster centre (\cref{fig:schematic_clustering}c).
\end{enumerate}
%
In this procedure, the critical parameter is the distance threshold $d_{max}$. This value balances the cluster size and partition similarity with the rest of the cluster. In this analysis, we use $d_{max}=0.10$,
implying that the best-matching communities' weighted average fraction of shared nodes is at least 90 percent.

\begin{figure}
    \begin{overpic}[width=\linewidth,tics=10]{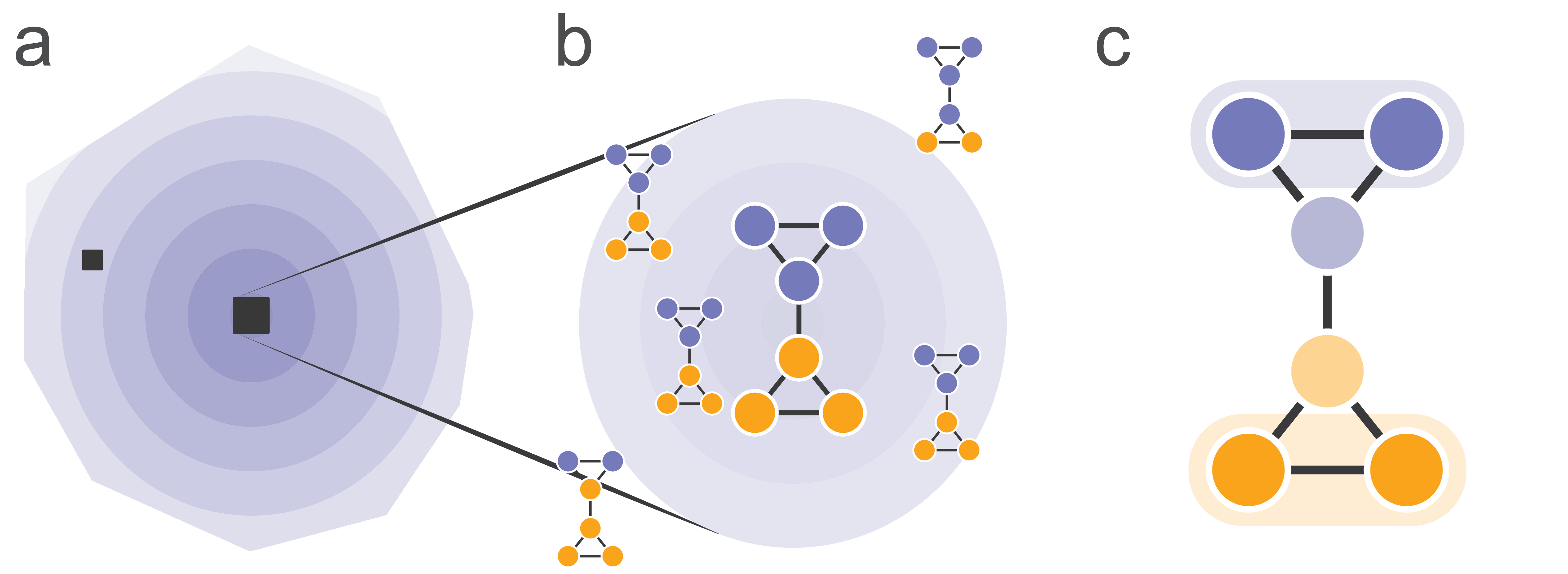}
    \put (10,15) {\large{$X$}}
    \put (52.3,14.8) {\large{$P$}}
    \put (93,10) {\textcolor[HTML]{FAA41C}{\large{$C_i$}}}
    \put (82.5,3) {$C_i'$}
    \put (39,2.2) {\footnotesize{$P_1$}}
    \put (41.5,22.5) {\footnotesize{$P_2$}}
    \put (44.5,13) {\footnotesize{$P_3$}}
    \put (61,10) {\footnotesize{$P_4$}}
    \put (61,29.5) {\footnotesize{$P_k$}}
    \end{overpic}
    \caption{Identifying core communities in a cluster centre.
    (a) The best-quality cluster $X$ (large~$\blacksquare$) in the solution landscape.
    (b) The cluster centre $P$ and co-clustered partitions $P_1,P_2,\dots,P_k$ inside the cluster with possibly different community assignments. 
    (c) Core communities of the best partition $P$ are found in a fraction~$p$ of the co-clustered partitions.}
    \label{fig:schematic_core_comm}
\end{figure}

\begin{figure*}[htp!]
    \centering
    \includegraphics[width=\linewidth]{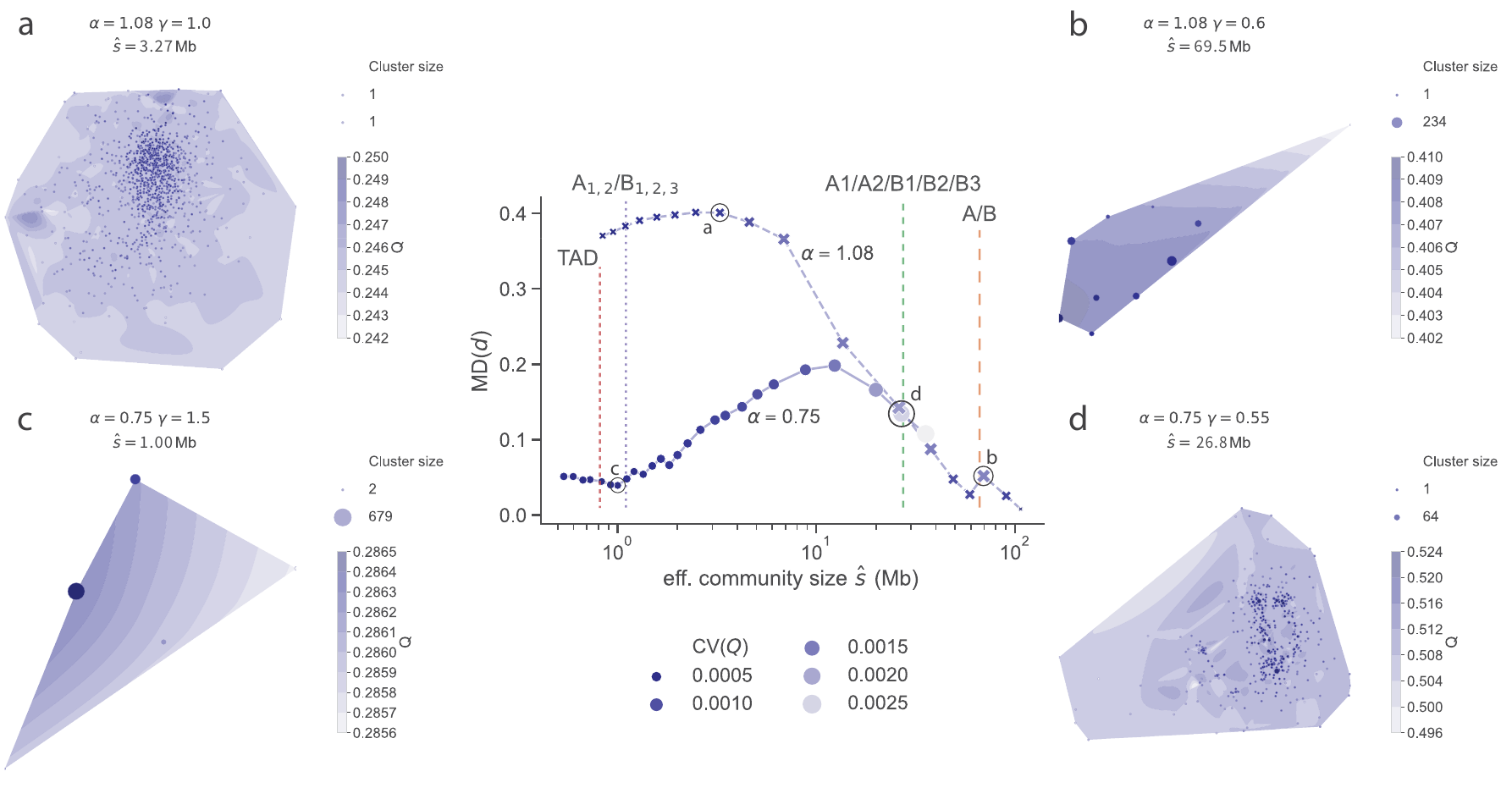}
    \caption{Solution landscapes at different scales. Mean difference (MD) of pairwise partition distances~$d$ for different $\alpha$ (main panel), surrounded by selected solution landscapes (a-d). The marker radius and colour are proportional to the quality $Q$'s coefficient of variation. As vertical lines, we show the effective sizes of chromatin divisions (summarised in \cref{tab:tad_ab_sizes}). TADs' effective size is 0.33~Mb. To fit them in the size axis, we show their effective size when omitting TADs smaller than five Hi-C-bins ($\sim 0.83$~Mb). We visualise the solution landscapes using DensMAP\cite{narayan2020density} on a contour plot of the quality scores. In panels a--d the distance between any two points is at least $d_{max}$.}
    \label{fig:mean_diff_chr10}
\end{figure*}

Next, after finding the cluster centres, we study if some network communities are more robust than others. 
We want to know if specific nodes co-appear in the same community in most partitions within a cluster while other nodes tend to change community memberships.
To do this, we first select clusters in the solution landscape with at least 100 partitions (\cref{fig:schematic_core_comm}a--b).
Then, we search for the largest node subset~$C_i'$ of each community~$C_i$ in $P$ that is clustered together in at least a fraction $p$ of the other co-clustered partitions.\cite{rosvall2010mapping}
We call these subsets core communities of the cluster centre (\cref{fig:schematic_core_comm}c).
The parameter $p$ balances core communities' size with how many partitions in the cluster that supports them.
We use $p=0.9$ to compensate for that the partitions in the clusters are allowed to differ by 90 percent on average.


\subsection{Mapping the solution landscape of human chromosome~10}

In this section, we study the degeneracy of the Hi-C network for human chromosome 10, applying the results from the previous section (see \cref{sec:data} for data handling).
Particularly, we wish to know how the solution landscape and core communities change with the parameter $\alpha$ associated with chromatin folding and GenLouvain's scale parameter $\gamma$ that sets the typical community size (see \cref{sec:louvain}). To make the ensuing discussion less abstract, we express $\gamma$ as a characteristic community size $\hat s$ (number of base pairs). This change simplifies the analysis, particularly when relating our results to established chromatin divisions.

Since the community sizes are relatively heterogeneous for most $\gamma$ values, we calculate $\hat{s}$ using the perplexity of the community sizes (see \cref{eq:eff-size,eq:eff-size-entropy} in \cref{sec:eff-size}).
We choose this metric because it is a better representation of characteristic sizes than the median or the arithmetic mean.
We depict the explicit $\hat s$--$\gamma$ relationships in \cref{fig:gamma-eff-size} for $\alpha = 0.75$ and $\alpha = 1.08$.

In \cref{fig:mean_diff_chr10}a--d, we plot the solution landscapes for four pairs of $\alpha$ and $\hat s$, each landscape spanning 1,000 GenLouvain runs.
Just as in \cref{fig:schematic_clustering,fig:schematic_core_comm}, we illustrate clusters as markers on top of $Q$ contour plots made using DensMAP.\cite{narayan2020density}
Each marker's diameter is proportional to the size of the cluster, and the colour represents the cluster's quality. 

The panels (a--d) illustrate typical landscape behaviours.
For example, (a) highlights a case where it is hard to find the optimal partition and distinguish the best community division because
all partitions have nearly identical qualities but have dissimilar community structures.
This leads to numerous size-one cluster centres scattered across the landscape.
As pointed out in \cref{tab:sl-classification}, we characterise this case as degenerate because there is substantial variability among the cluster centres pairwise distances and low variability in quality (high Var$(d)$ and low Var($Q$)).
So, in this case, we cannot be sure which cluster centre GenLouvain will gravitate towards from some random initial condition.

For larger community sizes ($\hat s \sim 70$ Mb), the solution landscape becomes much easier to analyse because we have only a few large clusters.
For example, in (b), GenLouvain recovers the same cluster centre most of the time.
Also, around (b), we find the most peaked solution landscapes where all partitions belong to a single cluster.

In panels (a) and (b), we used the looping exponent $\alpha  = 1.08$, which is the genome-wide averaged contact decay in human cells for distances $\gtrsim 1$ Mb.
However, $\alpha  = 0.75$ fits the data better for shorter distances (0.5--1.2~Mb).
With this in mind, we made similar analyses as above but for $\alpha = 0.75$ (\cref{fig:mean_diff_chr10}c--d).
This change made a noteworthy difference for the small communities [panel (c)]: the landscape has a clear cluster centre and a reliable, optimal solution.
However, forcing GenLouvain to assemble large communities with $\alpha = 0.75$ makes it increasingly degenerate up to a point (d) when the solution landscape has a global maximum alongside many local maxima with slightly lower $Q$.

Apart from these four examples, we made a parameter sweep of community sizes $\hat s$ for $\alpha = 0.75$ and $\alpha = 1.08$.
But instead of creating landscape plots for each parameter pair, we calculated the Jaccard distances $d_1, d_2,\ldots,d_i,\ldots$ [\cref{eq:jaccard-distance}] between all partition pairs. Then we calculated the simple average MD$(d) = E\left[d_i\right]$ and the coefficient of variation CV$(Q)$ of all partition qualities $Q_1, Q_2, \ldots$ .
The middle panel shows how MD$(d)$ varies with $\hat s$ for $\alpha = 1.08$ (crosses) and $\alpha = 0.75$ (circles) where we colour-coded the markers using CV$(Q)$.
This plot allows us to identify scale regimes where MD$(d)$ is large but CV$(Q)$ is small, which is a hallmark of a degenerate solution landscape.
For example, the plot demonstrates that $\alpha = 1.08$ is not a suitable folding parameter to find reproducible small-scale communities in the range  $\sim 1$--4~Mb.


In the middle panel, we also indicate $\hat s$ of published chromatin divisions, like TADs ($>0.5$~Mb) and A/B compartments (see \cref{sec:data}), by vertical dashed and dotted lines.
The scales close to (b) (encircled) corresponds to characteristic A/B compartment sizes, $\hat s = 66$~Mb.
Using $\alpha=1.08$, this scale is associated with a non-degenerate landscape leading to a reliable partition of the Hi-C network.
But interestingly, we note that there seems to be an even better division at a slightly smaller $\hat s$.
This panel also shows that we must use $\alpha = 0.75$ to find reliable partitions with sizes similar to TADs $\hat s = 0.33$~Mb.
Finally, sandwiched between A/B compartments and TADs, there is yet another commonly used Hi-C division denoted A1, A2, and B1,..., B3. This regime has less reliable communities because the landscape is flatter (exemplified in (d)). 

\subsection{Robust communities of chromosome~10}

\begin{figure}[htp]
    \centering
    \includegraphics[width=\linewidth]{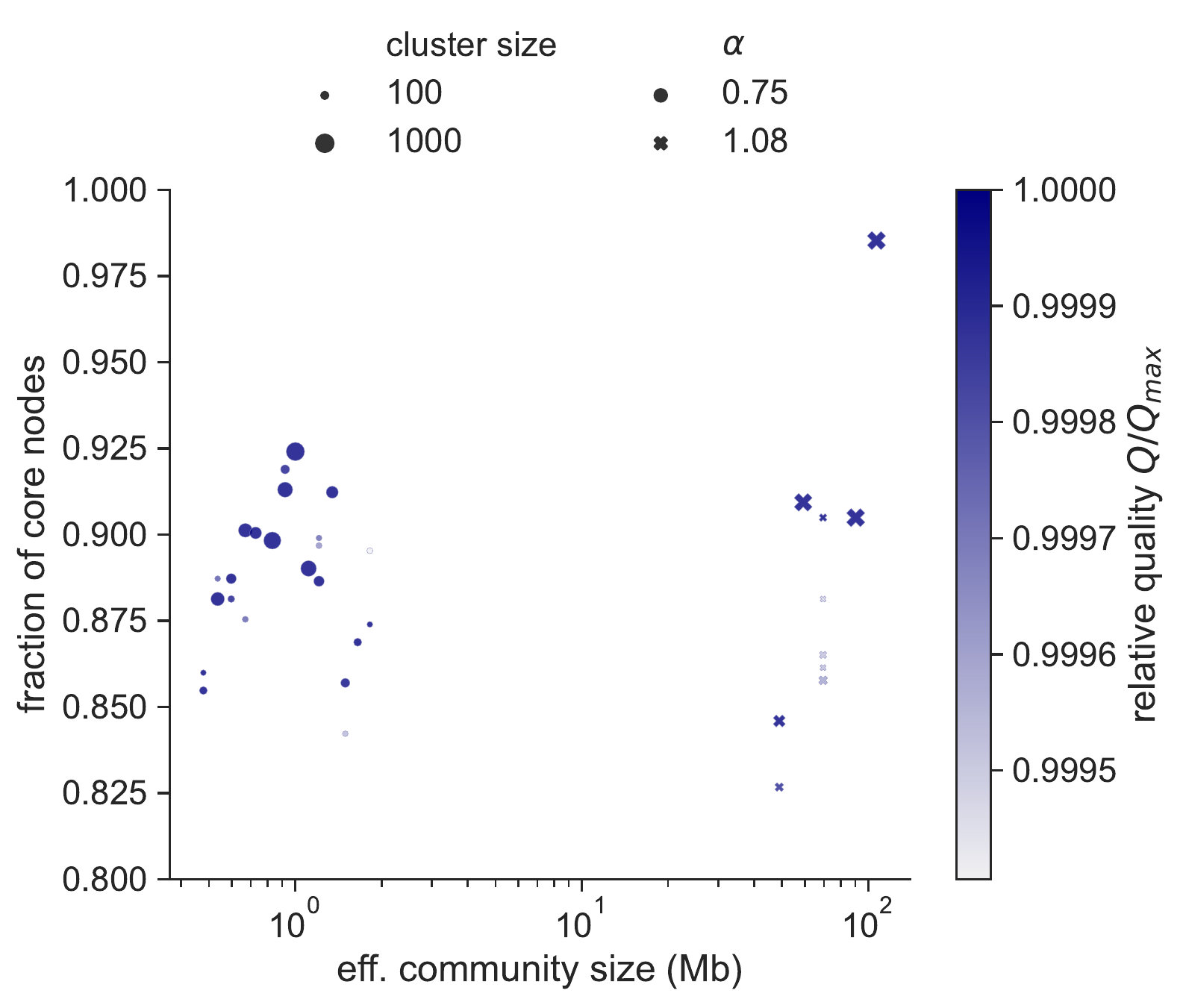}
    \caption{Cluster sizes and the fraction of core nodes. Core nodes are clustered in at least 90 percent of the cluster's partitions. We only show clusters with at least 100 partitions.}
    \label{fig:fraction_core}
\end{figure}

After classifying the solution landscape in \cref{fig:mean_diff_chr10}, we analyzed how robust the partitions are by identifying the core communities across $\hat s$.
As illustrated in \cref{fig:schematic_core_comm}, we extract robust communities by first clustering similar partitions and then quantifying the internal cluster differences. We quantify these differences by calculating the fraction of identical node-community memberships.
We omit clusters with less than ten percent of the total partition ensemble for a given $\hat s$--$\alpha$ combination (100 out of a 1,000 partitions).
We find robust communities when large clusters have a high fraction of nodes assigned to core communities (note marker sizes in \cref{fig:fraction_core}).
This finding holds for both folding parameters, $\alpha = 0.75$ and $\alpha = 1.08$.
Conversely, we find a fuzzy community structure when small clusters have the same relative quality $Q / Q_{max}$ and a small fraction of core-assigned nodes.


For $\alpha=0.75$, we observe that the most robust scale is $\hat s \sim 10^0$~Mb.
Here, one dominating cluster contains more than half of all partitions in which the communities contain nodes interacting primarily over short distances.
These communities are mostly unbroken DNA sequences (\cref{fig:alluvial-chr10}a) similar to TADs.
But there are exceptions.
For instance, we find a few large communities that join nodes from linearly separated DNA segments.
We illustrate the complete scale-dependent node-community memberships in \cref{fig:alluvial-chr10}a.
This figure shows how the nodes redistribute between communities when $\hat s$ changes.
Apart from observing stable communities (e.g, the beginning of the chromosome), we note that the 3D folding is not perfectly hierarchical, in which smaller communities form larger and larger super-structures.
Albeit small, there are deviations that make the folding structure semi-nested.\cite{bernenko2022mapping} 


For $\alpha=1.08$, we detect more than 80~percent core nodes when $\hat s > 40$~Mb and the most robust scale for $\hat s \sim 100$~Mb.
But this scale is trivially robust as most nodes are in a giant community (\cref{fig:alluvial-chr10}b). 
A more interesting case is where $\hat s \sim 60$~Mb and $\hat s \sim 90$~Mb, with the former having a slightly larger fraction of core node-assignments.
While $\hat s \sim 60$~Mb is similar to typical sizes of A/B compartments (\cref{fig:mean_diff_chr10}), we find multiple clusters when $\hat s \sim 70$~Mb that have similar quality but with lower core-node fractions.

\begin{figure}[htp!]
    \centering
    \includegraphics[width=\linewidth]{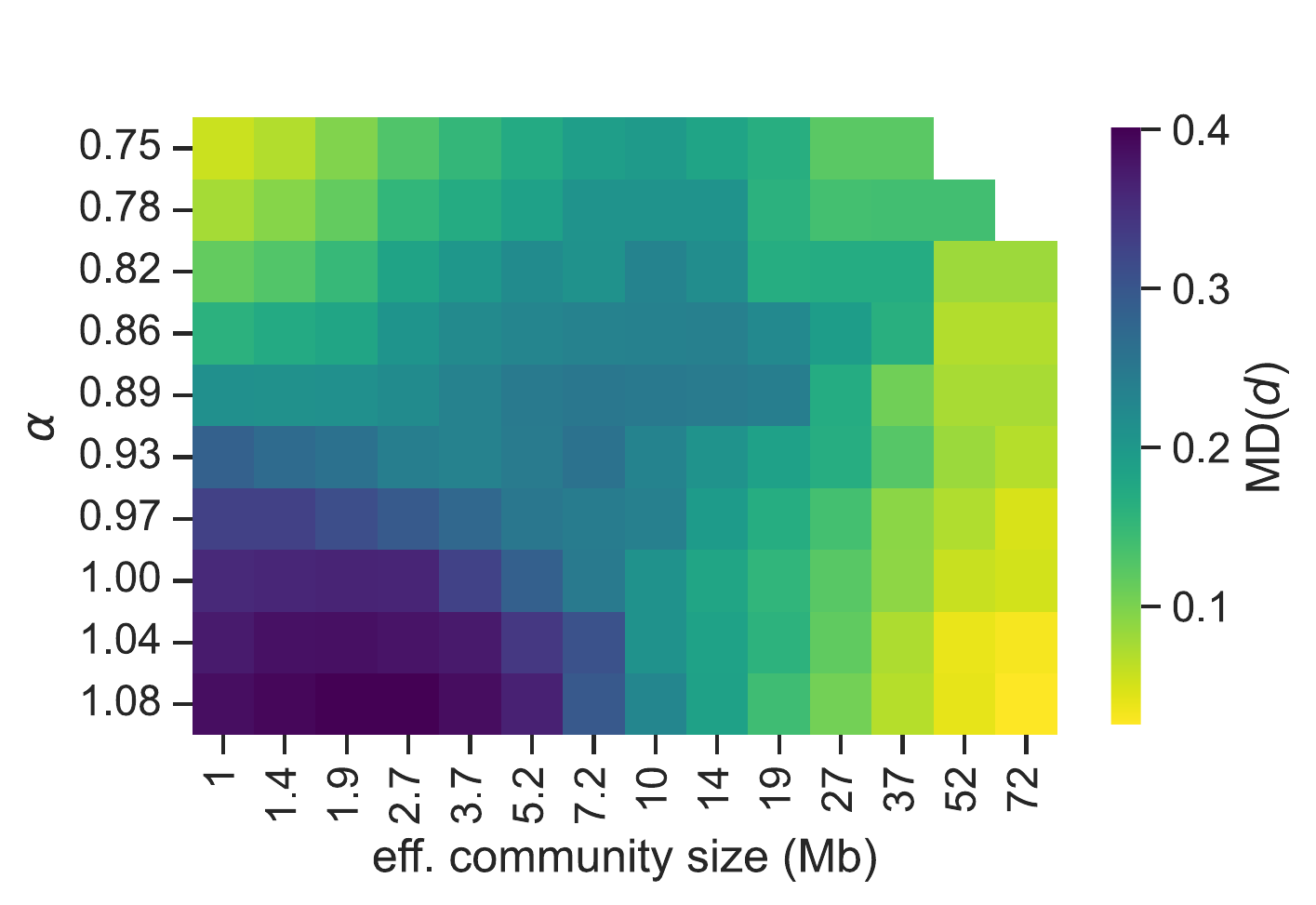}
    \caption{Mean absolute difference (MD) of pairwise partition distances~$d$ for chromosome~10 for different~$\alpha$.}
    \label{fig:eff_md_heatmap}
\end{figure}
Overall, we note that GenLouvain can detect reliable core communities at two distinct network scales ($\hat s \sim 1$~Mb and $\hat s \sim 60$~Mb) depending on the value of the folding parameter $\alpha$.
To investigate if there are other stable network scales, we made a sweep of $\alpha$ values for each $\hat s$ and calculated the mean partition distances MD$(d)$.
As shown in the heat map \cref{fig:eff_md_heatmap}, the most robust regimes are the top-left and bottom-right where MD$(d)$ is the smallest. 
In the bottom left corner where $\alpha \sim 1$ and $\hat s$ is small, we find the most degenerate solution landscape.

\subsection{Established chromatin divisions differ from optimal network communities}

In \cref{fig:mean_diff_chr10}, we indicated typical sizes of a few established chromatin divisions, like large TADs and A/B compartments, by vertical lines.
These chromatin divisions have size distributions that differ from typical network communities.
To make a better comparison, we varied $\gamma$ to find the network partition that is most similar to the chromatin divisions, disregarding that the effective size $\hat s$ may differ from $\hat s_{\rm TAD}$ or $\hat s_{\rm A/B}$.
Then we quantified the similarity 
by calculating the adjusted mutual information (AMI), commonly used to compare partitions.
The AMI is 1 when the two partitions are identical and 0 when inseparable from chance.
We summarise the results of our AMI analysis in \cref{tab:tad_ab_sizes}. 
\begin{table}[htp!]
\caption{Comparing optimal network partitions with established chromatin divisions. We derived the sizes for A1,A2,B1,...,B3 by aggregating A$_{1,2}$/B$_{1,2,3}$ segments ("A/B segments") and the A/B sizes from merging A1/\dots/B3 sub-compartments (see \cref{fig:ab-scales}). Notation: effective size~$\hat{s}$, and adjusted mutual information (AMI), chromatin looping exponent $\alpha$. We found no similar partition for A1,\dots,B3.}
\label{tab:tad_ab_sizes}
{\setlength{\tabcolsep}{0.5em}
\sffamily
\begin{tabular}{@{}lS[table-format=2.2]S[table-format=2.1]@{\hskip 20pt }S[table-format=1.2]S[table-format=2.2]S[table-format=1.2]@{}}
                        &       & \\
                        & \multicolumn{2}{c}{Characteristic} & \multicolumn{3}{c}{Most similar} \\
                        & \multicolumn{2}{c}{size (Mb)} & \multicolumn{3}{c}{partition} \\ \addlinespace[2pt]
                        & \multicolumn{1}{c}{median} & \multicolumn{1}{c}{$\hat{s}\quad$} & \multicolumn{1}{c}{$\alpha$} & \multicolumn{1}{c}{$\hat{s}$} & AMI \\ \addlinespace[3pt]
TADs                & 0.18   & 0.33   & 0.75  & 0.77  & 0.53 \\
A/B segments   & 0.30  & 1.1   & 0.75  & 1.8   & 0.72 \\
A1,...,B3          & 31    & 27    & {---} & {---} & {---} \\
A/B compartments                  & 64    & 66    & 1.08  & 59    & 0.47      
\end{tabular}}
\end{table}

For TADs (\cref{tab:tad_ab_sizes}, top row), we find the best correspondence when $\hat s = 0.77$~Mb, which is larger than TADs' effective size $\hat s_{\rm TAD} = 0.33$~Mb.
Here, the AMI score is 0.53, indicating that the community structures show significant deviations.
This deviation is likely because median TAD sizes are close to the data resolution we use (0.1~Mb).
The AMI score is similar for A/B compartments (AMI $= 0.47$), but the scales match better ($\hat s = 66$~Mb vs $\hat s = 59$~Mb).
We find the best overlap with the small-scale A$_{1,2}$/B$_{1,2,3}$ segments (denoted ``A/B segments'' in \cref{tab:tad_ab_sizes}) with  $\hat s = 1.8$~Mb and AMI $=0.72$.
We do not compare our results with A1/A2/B1/B2/B3 sub-compartments because we cannot detect robust communities in this regime.

\begin{figure}[hpt!]
    \centering
    \includegraphics[width=0.9\linewidth]{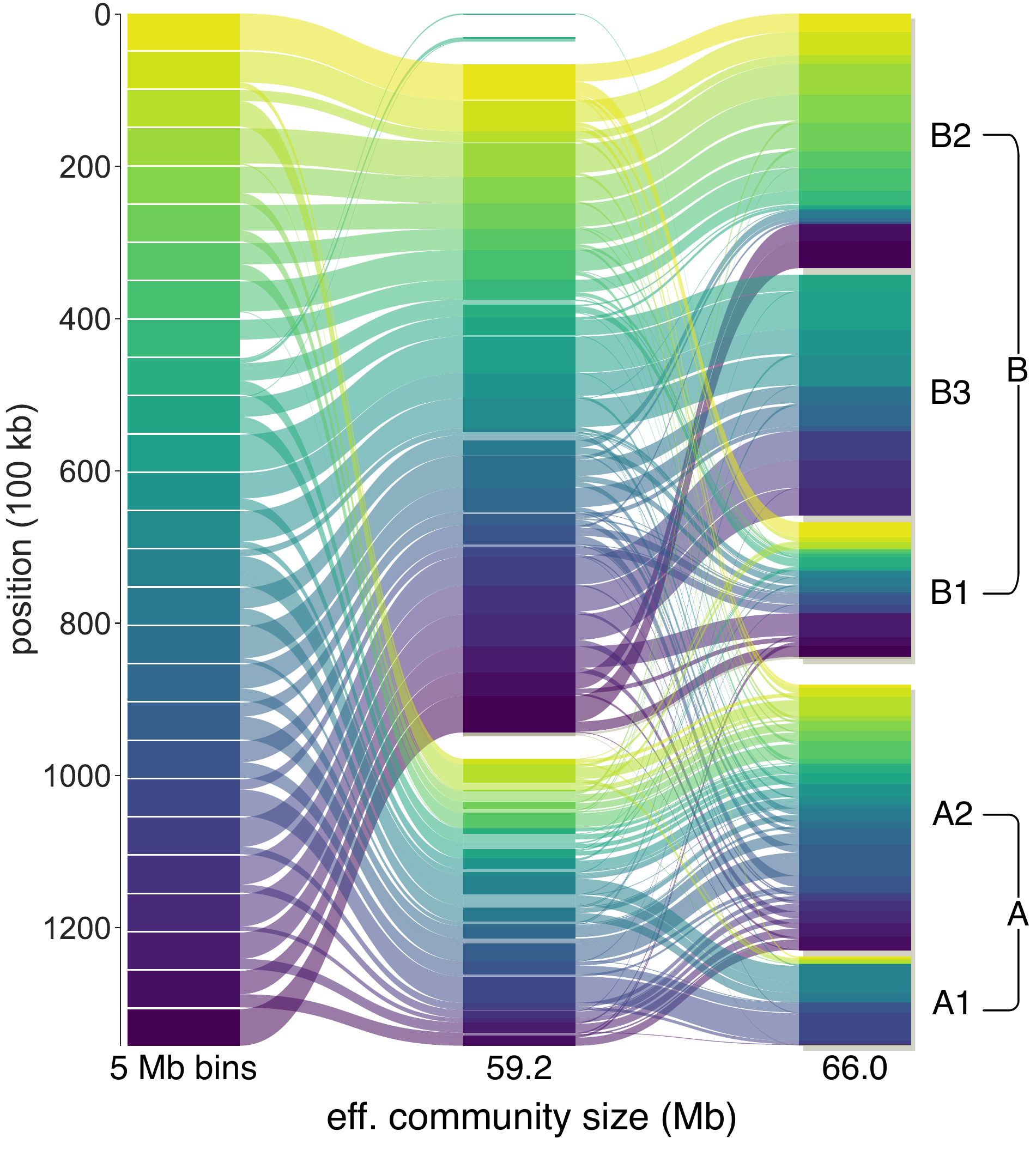}
    \caption{Core communities and A/B compartments of chromosome~10. The leftmost column represents 5~Mb bins coloured by position. The middle column represents the partition most similar to the A/B compartments ($\hat s = 66$~Mb), with communities ordered by average position. Transparent segments do not belong to the core. The right-most column represents A/B sub-compartments.}
    \label{fig:a12b123}
\end{figure}

Finally, in \cref{fig:a12b123}, we visualise how the node-community membership differs between the A/B compartments and the optimal network partition at $\hat s = 59$~Mb.
We observe that most sub-compartments are isolated into a single network community.
But the A2 sub-compartment includes Hi-C bins assigned to the two largest communities.

%
\section{Discussion}

Hi-C networks are densely connected.
Therefore, finding reliable community structures across various scales is challenging.
To better understand this problem, we have mapped out the solution landscape of feasible partitions in a chromosome contact network at different organization scales.
We sampled 1,000~partitions using different scale- and DNA-looping parameters to detect regimes associated with robust or degenerate solution landscapes.
We classified these regimes in terms of the variabilities of the partition's qualities and pairwise distances. Then we used a partition clustering approach and compared cluster sizes and qualities.
Also, studying the proximity of the best-quality partition, we find robust core communities supported by at least 90~percent of the proximate partitions.
Finally, varying the looping parameter $\alpha$ We find robust small-scale communities for $\alpha=0.75$ and larger-scale communities for $\alpha=1.08$, roughly corresponding to TADs and A/B compartments.
Between these extremes, we find a regime opaque to community detection methods.
 
We mapped out the multiscale solution landscape in \cref{fig:mean_diff_chr10} and discovered regimes where the landscape is degenerate, as illustrated in panel (a).
It is critical to note this degeneracy problem is not easily resolved using another community detection method because there might not exist strong communities in the data at that scale.
Therefore, different methods will provide different answers.
We circumvented some degenerate scales by modifying the null model's folding parameter.
For example, at $\hat s \sim 1$~Mb, changing $\alpha$ from 1.08 to 0.75, GenLouvain recovers the same optimal partition most of the time. However, this approach is not straightforward to generalise.

Furthermore, we found two distinct regimes in the $\alpha$--$\hat s$ parameter space where community detection is easy (in \cref{fig:eff_md_heatmap}).
But this finding does not exclude other robust network scales.
In GenLouvain's modularity function, we assumed that node-node contacts decay as a power law with some exponent $\alpha$.
While this is consistent with the average contact decay in Hi-C maps and established polymer physics models (e.g., the Gaussian chain or the fractal globule), there could be other functional forms that better describe the actual folding mechanism or a blend of several competing mechanisms (e.g., short-ranged loop-extrusion and long-ranged phase separation).\cite{nuebler2018chromatin}
This amounts to improving the null model, which we leave as an avenue for future research.

We found that established chromatin divisions differ from the optimal GenLouvain partition associated with identical characteristic sizes (\cref{tab:tad_ab_sizes}). 
Even if sweeping through a range of characteristic sizes, we still find significant differences with the most similar GenLouvain partition.
We achieved the best match for A$_{1,2}$/B$_{1,2,3}$ segments, and the matching communities are robust.
While we cannot reach perfect overlap using one single characteristic size, we point out that it is conceivable to increase the overlap if considering partitions from several $\hat s$.
This indicates that our approach might find most chromatin divisions but not at a single $\hat s$.
This finding helps benchmark our results to other published TAD-finding methods and offers a systematic approach to highlight deviations from expected network partitions under the null model (power law decaying contacts).

While this work focuses on Hi-C contact maps, GenLouvain is commonly used to detect communities in a wide range of networks. Therefore, our work is helpful to other researchers searching for robust communities
when facing the degeneracy problem. 



\section{Materials and methods}\label{sec:methods}

\subsection{Assembling chromosome contact data}\label{sec:data}

We downloaded Hi-C data for the B-lymphoblastoid human cell line (GM12878)\cite{rao20143d} from the GEO database (\texttt{MAPQG0} dataset, 100~kb resolution).\cite{edgar2002gene}  
The data file contains measured contact frequencies between DNA segment pairs in a cell population. 
We only consider intra-chromosome contacts in our analysis, allowing us to study each chromosome by itself. 
We interpret the Hi-C data  as a weighted network in sparse form, where each node represents a 100~kb DNA segment, and the link weight is the measured contact count. 
Before constructing the network, we normalise the data using the Knight-Ruiz matrix balancing algorithm. 

In addition to Hi-C data, we use datasets associated with existing 3D divisions:\cite{rao20143d} A/B sub-compartments and topologically associating domains (TAD) (downloaded from the GEO database \cite{edgar2002gene}). The sub-compartments divide chromosomes into regions called A1, A2, B1, B2, B3, and B4. While A1 and A2 exhibit high gene expression, B1--B3 are associated with repressed and inactive DNA regions (B4 is found only in chromosome~19 and does not participate in our study as we focus on chromosome~10). Also, functionally similar sub-compartments tend to have correlated contact patterns and are generally referred to as A- and B-compartments.
Alongside the sub-compartment, we study TADs. Defined by the Arrowhead algorithm,\cite{rao20143d} TADs are genomic regions with above-average contact frequencies, serving as microenvironments for co-regulated genes. TADs appear as squares along the main diagonal in Hi-C maps.

\subsection{Multiscale community detection}\label{sec:louvain}

To find network communities, we use the Generalized Louvain method (GenLouvain).\cite{jeubgeneralized} GenLouvain searches for network partitions that maximise the modularity function $Q$, capturing local deviations from the expected background connectivity. While the most common choice is random connections, better known as the Newman-Girvan null model,\cite{newman2004finding} we rescale the expected link weights to mimic that nodes are interconnected DNA segments forming a long polymer chain that is folded in 3D inside the cell nucleus.\cite{lee2019mapping} Empirical data shows that the average link weight ($\propto$ number of contacts) decays as a power-law with linear node separation.  After this modification, the parametric modularity (or quality) function is\cite{reichardt2006statistical}
\begin{equation}
    \label{eq:modularity}
    \begin{split}
         Q &= \frac{1}{2m} \sum_{i \neq j} \left( A_{ij} - \gamma P_{ij}^{(0)} \right) \delta(C^i, C^j),  \\
         &P_{ij}^{(0)}= \frac{ 2m \, k_i \, k_j \, |i - j|^{-\alpha} }{ \sum_{i' \neq j'} k_{i'} \, k_{j'} \, |i' - j'|^{-\alpha} },
    \end{split}
\end{equation}
where  $A_{ij}$ are entries in the weighted adjacency (Hi-C) matrix, $m$ is the total weight, $\gamma$ is the scale parameter, $k_i$ is the strength of node~$i$, and $C^i$ is node $i$'s community assignment.
By tuning the scale parameter $\gamma$, we get a spectrum of communities of different sizes. With increasing $\gamma$, we penalise any links with weights close to the random expectation.

The decay parameter $\alpha$ reflects DNA's 3D folding. This parameter also changes how GenLouvain treats weak (or long-ranged) connections when assembling communities. Particularly, while decreasing~$\alpha$ tend to disfavour weak links, working as a threshold for long-range links, increasing $\alpha$ favour weak links. When $\alpha=0$, we recover the Newman-Girvan null model.
Based on empirical data, we study~$\alpha=1.08$ to find large, long-range ($\sim 500$--7000~kb) communities,\cite{lieberman2009comprehensive} and~$\alpha=0.75$ to find smaller, short-range ($\sim 200$--1200 kb) communities.\cite{sanborn2015chromatin}
These values reflect two DNA-folding mechanisms: the loop extrusion that forms small-scale 3D structures, and the phase separation that governs the self-aggregation of distant regions.

Finally, we set GenLouvain to randomly regroup nodes to communities proportional to the resulting quality increase. This achieves better solution landscape sampling.

\subsection{Characteristic community size}\label{sec:eff-size}

We explore the solution landscapes over varying scale and decay parameters. To compare the partitions' characteristic community sizes, we use a metric that is weakly dependent on spurious singleton communities, unlike the mean and median. Instead, we use the effective community size
\begin{equation}
    \label{eq:eff-size}
    \hat{s} = \frac{\text{number of nodes}}{\text{effective number of communities}},
\end{equation}
where we calculate the effective number of communities using the perplexity $2^{H(P)}$ of partition~$P$'s community size distribution, with Shannon entropy
\begin{equation}
    \label{eq:eff-size-entropy}
    H(P) = - \sum_i \frac{|C_i|}{\sum_j |C_j|} \log_2 \frac{|C_i|}{\sum_j |C_j|}.
\end{equation}

\section{Data availability}

The \texttt{MAPQG0} dataset, sub-compartment, and topologically associating domain (TAD) data\cite{rao20143d} was downloaded from The GEO Database at \url{https://www.ncbi.nlm.nih.gov/geo/}.

\section{Author contributions}

A.H.\ and L.L.\ devised the study.
D.B.\ prepared the data.
A.H.\ and D.B.\ performed the experiments and analysed the results.
All authors wrote, edited, and accepted the manuscript in its final form.

\section{Competing interests}

The authors declare that they have no competing interests.

\acknowledgements

The authors would like to thank Martin Rosvall, Magnus Neuman, and Jelena Smiljanić for the feedback that improved this manuscript.
A.H.\ was supported by the Swedish Foundation for Strategic Research, Grant No. SB16-0089.

\bibliographystyle{unsrt}
\bibliography{main}

\appendix
\clearpage
\section*{Supplementary Information}\label{sec:SI}

\setcounter{figure}{0}
\renewcommand{\thefigure}{S\arabic{figure}}

\begin{figure}[htp]
    \centering
    \includegraphics[width=0.9\linewidth]{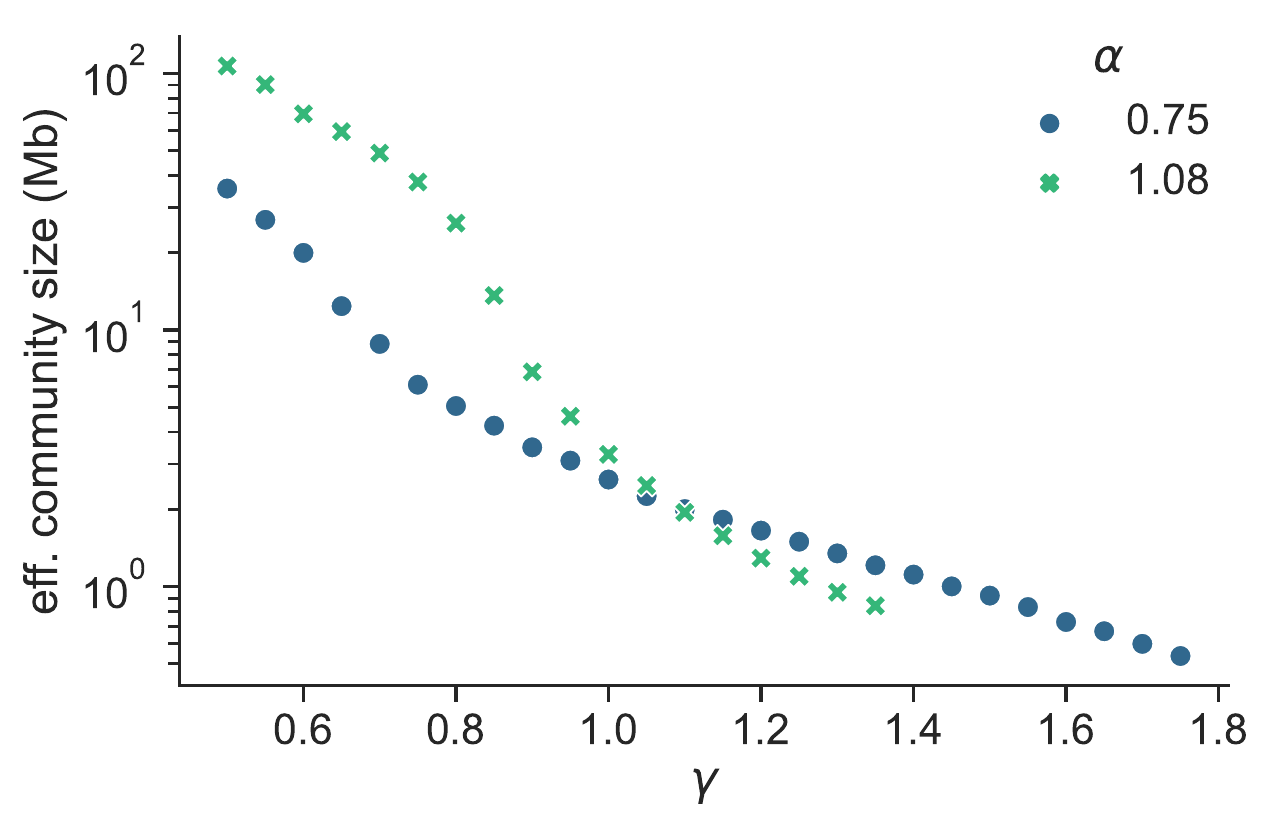}
    \caption{Effective community size for different scale parameters $\gamma$ and decay parameters $\alpha$ for chromosome~10.}
    \label{fig:gamma-eff-size}
\end{figure}

\begin{figure}[htp]
    \centering
    \includegraphics[width=\linewidth]{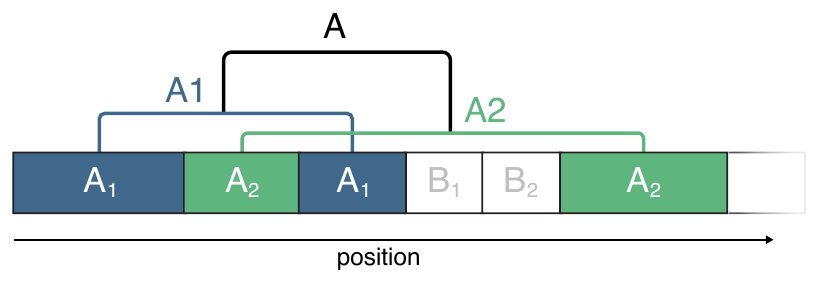}
    \caption{Schematic structural scales derived from A1/A2/B1/B2/B3 sub-compartment data.\cite{rao20143d}
    The smallest organisational scale is A$_{1,2}$/B$_{1,2,3}$-segments, a contiguous DNA stretch that fully belongs to a sub-compartment. The next scale is A1/\dots/B3 sub-compartments -- collections of DNA segments. The largest structural scale is the A/B compartments.}
    \label{fig:ab-scales}
\end{figure}

\begin{figure*}[htbp]
    \centering
    \includegraphics[width=\linewidth]{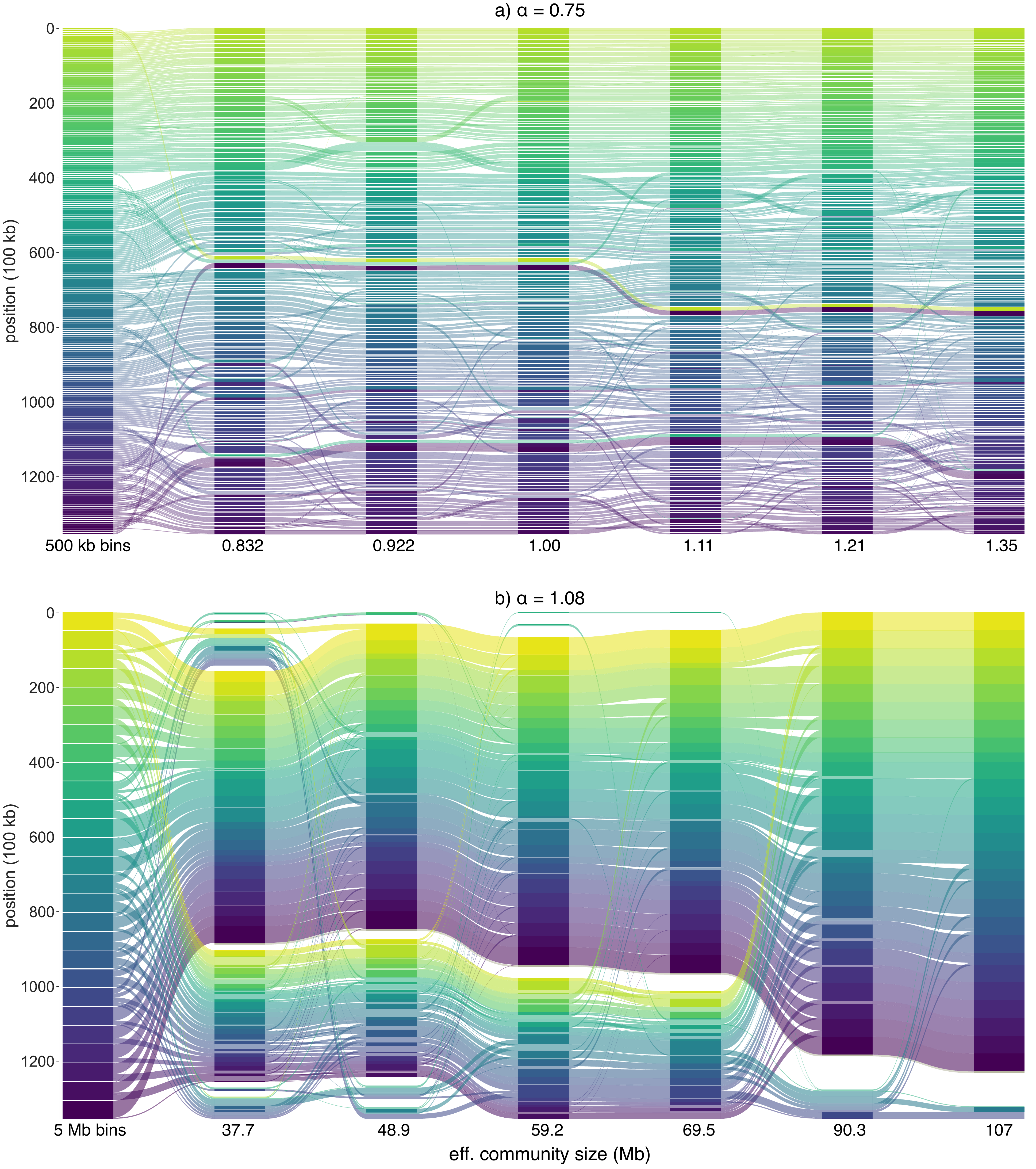}
    \caption{Alluvial diagram of core communities of chromosome~10 for $\alpha=0.75$ and $\alpha=1.08$ at different scales. The left-most column represents linear bins coloured by position. The remaining columns represent community structure at different scales, vertically ordered by their average position and coloured by the positions of their contained segments. Transparent segments are not in the core.}
    \label{fig:alluvial-chr10}
\end{figure*}

\end{document}